\documentclass[twocolumn,preprintnumbers,amsmath,amssymb]{revtex4}

\usepackage{graphicx}
\usepackage{dcolumn}
\usepackage{bm}

\begin{document}

\title{Inelastic collisions of ultra-cold heteronuclear molecules in an optical trap}

\author{Eric R. Hudson}
\email{Eric.Hudson@Yale.edu}
\author{Nathan B. Gilfoy}
\author{S. Kotochigova}
\altaffiliation{Department of Physics, Temple University,
Philadelphia, Pennsylvania 19122-6082, USA}
\author{Jeremy M. Sage}
\altaffiliation{Lincoln Laboratory, Massachusetts Institute of
Technology, Lexington, MA, 02420, USA}
\author{D. DeMille}
\affiliation{Department of Physics, Yale University, 217 Prospect
Street, New Haven, CT 06511, USA}

\begin{abstract}
Ultra-cold RbCs molecules in high-lying vibrational levels of the
a$^3\Sigma^+$ ground electronic state are confined in an optical
trap. Inelastic collision rates of these molecules with both Rb and
Cs atoms are determined for individual vibrational levels, across an
order of magnitude of binding energies. A simple model for the
collision process is shown to accurately reproduce the observed
scattering rates.
\end{abstract}

\maketitle

The electric dipole-dipole interaction provides a long-range,
tunable anisotropic interaction between polar molecules. This is
fundamentally different from most interactions studied between
ultra-cold atoms, which are typically isotropic and comparatively
short-ranged. Features of the dipole-dipole interaction can lead to
many novel and exciting phenomena, such as field-linked states
\cite{avdeenkov_field-linked_2004}, long-range topological order
\cite{micheli_toolbox_2006}, quantum chemistry
\cite{roman_v_krems_molecules_2005,hudson_production_2006}, and the
possibility for quantum computation \cite{demille_quantum_2002,
rabl_hybrid_2006}. Furthermore, the presence of closely spaced
internal levels of the molecules, \textit{e.g.} $\Omega$-doublet,
rotational, and vibrational levels, presents a host of new
possibilities for precision measurement of fundamental physics
\cite{hudson_cold_2006,flambaum_enhanced_2007,demille_using_2008,
demille_[07090963v1]_,zelevinsky_[07081806]_}. Producing
\textit{ultra-cold} samples of polar molecules will facilitate
trapping and, thus, the required high densities and long observation
times for observing these phenomena.

Techniques such as Stark deceleration
\cite{bethlem_decelerating_1999} and buffer gas cooling
\cite{doyle_buffer-gas_1995} are capable of producing cold samples
from a wide range of molecular species; however, the temperatures
and densities currently attainable via these ``direct cooling''
methods are not sufficient for observing many of the interesting
phenomena mentioned above. Conversely, the association of ultra-cold
atoms, either via a Feshbach \cite{inouye_observation_2004} or
optical resonance \cite{kerman_first_2004}, restricts experiments to
a limited class of molecules -- namely, those composed of laser
cooled atoms. Nonetheless, these methods are approaching
temperatures and densities appropriate for observing the
aforementioned phenomena.

\begin{figure}
\resizebox{1\columnwidth}{!}{
    \includegraphics{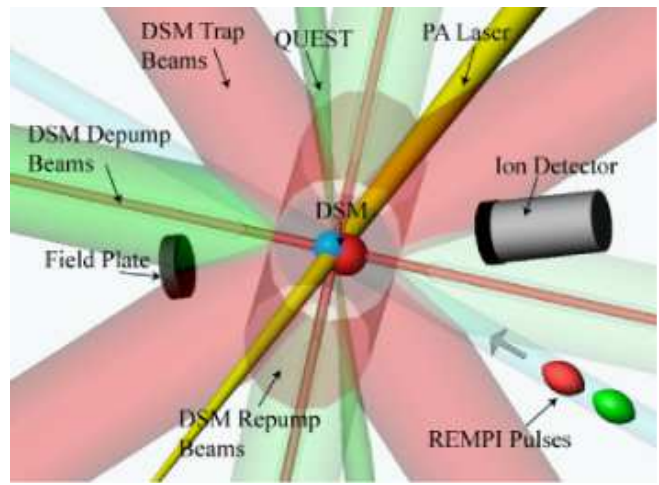}%
} \caption{(Color Online) Schematic of the experiment showing the
overlap of the DSM with the relevant beams and relative position of
the ion detector utilized in the state-selective REMPI
detection.\label{ChamberFigure}}
\end{figure}

In this Letter, we report the optical confinement of ultra-cold,
vibrationally excited RbCs molecules in the a$^3\Sigma^+$ ground
electronic state, produced via photo-association (PA) of
laser-cooled $^{85}$Rb and $^{133}$Cs atoms. We utilize the long
observation times afforded by the optical trap to determine the
inelastic scattering rate for specific vibrational levels of these
molecules, with both $^{85}$Rb and $^{133}$Cs atoms, across an order
of magnitude of binding energies. We show that a simple model for
the collision process accurately reproduces the observed scattering
rates. We also extend this model to estimate molecule-molecule
inelastic scattering rates and discuss implications for producing
trapped samples of X$^1\Sigma(v = 0$) RbCs molecules.

\begin{figure}
\resizebox{1\columnwidth}{!}{
    \includegraphics{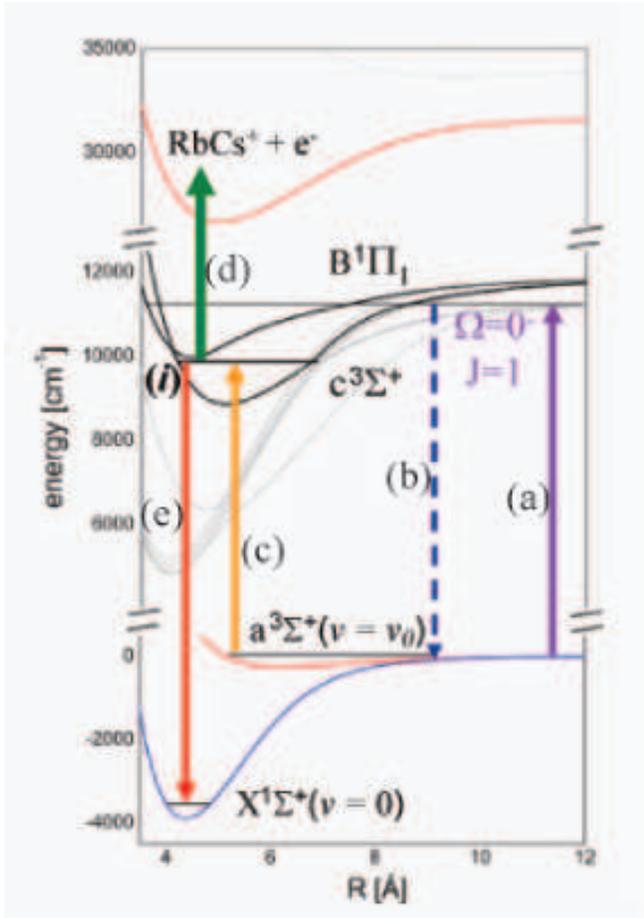}%
} \caption{(Color Online) Formation and detection processes for
ultracold RbCs. (a) The PA process excites colliding atom pairs into
bound RbCs$^*$ molecules, which (b) decay into a range of
vibrational states of the a$^3\Sigma$ potential. (c) Metastable
a$^3\Sigma(v)$ molecules are excited to level i, then (d) ionized
and subsequently detected via time-of-flight mass spectrometry. The
application of Stokes light (e) instead of the ionizing pulse (d)
can be used to produce molecules in the absolute ground state
X$^1\Sigma(v = 0)$. \label{RbCsPotentials}}
\end{figure}

The apparatus used in this work is shown in
Fig.~\ref{ChamberFigure}. Briefly, $^{85}$Rb and $^{133}$Cs atoms
are cooled and collected in a dual-species, forced dark-spot
magneto-optical trap (DSM)
\cite{anderson_reduction_1994,kerman_first_2004}. Using absorption
imaging along two orthogonal directions we co-locate the species and
measure the atomic density, $n$, and atom number, $N$, as $n_{Rb}$ =
4(2)$\times$10$^{11}$ cm$^{-3}$, $N_{Rb}$ = 9(1)$\times$10$^{7}$,
and $n_{Cs}$ = 5(1)$\times$10$^{11}$ cm$^{-3}$, $N_{Cs}$ =
2(1)$\times$10$^{8}$. The temperature, $T$, of each species in the
DSM was measured by time-of-flight expansion to be $T_{Rb}$ = (80
$\pm$ 25) $\mu$K and $T_{Cs}$ = (105 $\pm$ 40) $\mu$K. Our optical
trap is a quasi-electrostatic trap (QUEST) realized in a 1-D lattice
configuration. The QUEST, represented by the green beam in Fig.
\ref{ChamberFigure}, is formed by focusing and retro-reflecting the
beam from a vertically-aligned 100 W CO$_2$ laser, operating with a
10.6 $\mu$m wavelength. An acousto-optical modulator placed in the
beam path allows for the rapid turn-off of the QUEST ($\tau$ $<$ 1
$\mu$s) and serves as an optical isolator for light reflected back
into the laser. The $e^{-2}$ intensity beam waist at the focus is
$\sim$75 $\mu$m, yielding a peak intensity of $\sim$3 MW/cm$^2$ and
trap depths of $\sim$4 mK, $\sim$6 mK, and $\geq$9 mK for Rb, Cs,
and a$^3\Sigma^+$ RbCs, respectively. Because we utilize a lattice
configuration for our QUEST, we are not restricted to trapping at
the focus. We find it advantageous to trap atoms and molecules away
from the focus, where the trap volume is much larger. By moving the
point of overlap between the QUEST and DSM, we trap $\sim$9 mm away
from the focus, where the waist is $\sim$400 $\mu$m and the trap
depths are reduced by $\sim$30. In addition to providing a larger
trapping volume, this method mitigates the effects of QUEST-induced
light-shifts \cite{griffin_spatially_2006}.

The energy level scheme relevant to the RbCs formation is shown in
Fig. \ref{RbCsPotentials}. The PA laser has an intensity of $\sim$2
kW/cm$^2$, and its frequency is locked to an $\Omega$ = 0$^-$, $J_P$
= 1$^+$ level, located 38.02 cm$^{-1}$ below the Rb 5S$_{1/2}$(F= 2)
+ Cs 6P$_{1/2}$(F = 3) atomic asymptote \cite{kerman_first_2004}.
Spontaneous decay of this state primarily produces molecules in the
a$^3\Sigma^+$ state vibrational level with binding energy $E_B$ = -
5.0 $\pm$ 0.6 cm$^{-1}$, to which we assign vibrational number
$v_0$. In our previous work \cite{kerman_second_2004} we tentatively
assigned $v_0 = 37$, but this value has an uncertainty of several
units since the depth of the $a^3\Sigma^+$ state potential well is
not accurately known.

\begin{figure}
\resizebox{1\columnwidth}{!}{
    \includegraphics{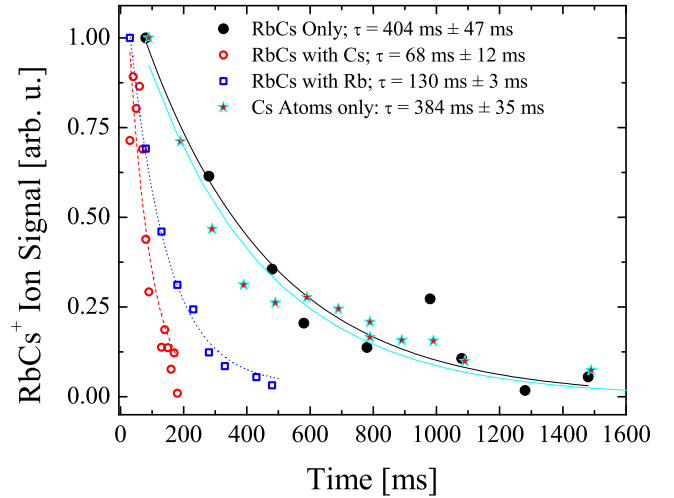}%
} \caption{(Color Online) Typical molecular lifetime data. Here the
number of molecules in the  a$^3\Sigma^+(v = v_0$) state with
binding energy $E_B$= - 5.0 $\pm$ 0.6 cm$^{-1}$ is observed in the
QUEST as a function of time. The presence of inelastic collisions
between the atoms and molecules is evidenced by the dramatic
reduction of the molecular lifetime when atoms are present. With no
atoms we observe molecule lifetimes consistent with the background
gas limited lifetime seen for isolated atomic clouds in the
trap.\label{LifetimeData}}
\end{figure}

Experimental data is taken by loading the DSM for 5~s from
background alkali vapor, provided by heated getters, in the presence
of both the PA laser and QUEST. Atoms are not efficiently loaded
into the lattice directly from the DSM due to the low trap depth
($\sim$100-200 $\mu$K) away from the focus. Because the QUEST is
substantially deeper for the molecules than the atoms ($\geq$300
$\mu$K), the PA process efficiently loads RbCs into the lattice.
Thus, molecules are loaded into the QUEST simply by applying the PA
beam during the DSM loading process. From the measured atomic
densities and known PA rates \cite{kerman_second_2004}, we estimate
that we trap $N_{RbCs}$ $\approx$ $10^{4}$ molecules at a density of
$n_{RbCs}$ $\approx$ $10^{9}$ cm$^{-3}$ and temperature of
$T_{RbCs}$ $\approx$ 100 $\mu$K with roughly 7\% in the a$^3\Sigma(v
= v_0$) state. To study atom-molecule collisions we load atoms into
the lattice more efficiently via an optical molasses cooling stage
for the desired atomic species, after the DSM is loaded. The optical
molasses stage is performed by shifting the detunings, $\Delta$, of
the DSM trap lasers to $\Delta_{Rb}$ = -6$\Gamma$, $\Delta_{Cs}$ =
-16$\Gamma$ for 10 ms, where $\Gamma$ is the transition natural
linewidth. While the hyperfine depumping beam of the DSM remains on
during the entire molasses stage, the DSM hyperfine re-pumping beam
is extinguished for the last 100 $\mu$s to ensure that all trapped
atoms are in the lowest (dark) hyperfine state. Loading the lattice
in this way leads to typical densities of $n_{Rb}$ =
2(1)$\times$10$^{11}$~cm$^{-3}$ and $n_{Cs}$ =
6(1)$\times$10$^{11}$~cm$^{-3}$, and temperatures of $T_{Rb}$ = (20
$\pm$ 11)~$\mu$K and $T_{Cs}$ = (20 $\pm$ 15)~$\mu$K in a volume of
$\approx\pi$(89 $\mu$m)$^2\times$~960 $\mu$m, which is roughly a
factor of 40 increase in density and factor of 5 reduction in
temperature compared to loading directly from the DSM. After the
molasses stage, we apply resonant `push'-beams for 10 ms to remove
any undesired atoms from the lattice.  After the `push'-beam
sequence, all beams except the QUEST are shuttered, and the
molecules and any deliberately trapped atoms are held in the
lattice.

After a variable delay time, the QUEST is switched off and the
trapped molecules are state-selectively ionized using
Resonance-Enhanced Multi-Photon Ionization (REMPI), as shown in Fig.
\ref{ChamberFigure}. The resulting ions are detected using
time-of-flight mass spectrometry \cite{sage_optical_2005}. In this
manner, we use the observed trap-lifetime of the molecules in the
QUEST as a direct measurement of the molecular collision rates.

Typical lifetime data is shown in Fig. \ref{LifetimeData} for
molecules in the  a$^3\Sigma^+(v = v_0$) state. As can be seen, the
presence of atoms in the lattice significantly shortens the lifetime
of the trapped molecules, which is otherwise limited by collisions
with background gas. We attribute this behavior to inelastic
collisions between the atoms and molecules. These losses are likely
due to ro-vibrational quenching or hyperfine changing collisions.
Each of these degrees of freedom carries sufficient energy that its
relaxation creates enough kinetic energy to remove both the molecule
and the atom from the trap, \textit{e.g.} one vibrational quantum is
$\sim$2 K. The number of trapped molecules, $N_{RbCs}$, evolves in
time according to:
\begin{eqnarray}
\frac{dN_{RbCs}}{dt} &=& -
\Gamma_{BG}N_{RbCs} \nonumber \\
 & & - \Gamma_{atom}N_{RbCs} - \frac{\beta}{V}
N_{RbCs}^2. \label{MoleculeNumberEqn}
\end{eqnarray}
Here $\Gamma_{BG}$ is the loss rate due to collisions with
background gas, $\Gamma_{atom}$ is the loss rate due to inelastic
collisions with atoms, $\beta$ is the molecular 2-body loss rate,
and $V$ is the trap volume occupied by the molecules. Since two-body
processes are negligible compared with background gas collisions
($\beta n_{RbCs}/\Gamma_{BG} \ll 1$), we use a fit of the data to
the form of $N_{RbCs}(t) = N_{o}e^{-t/\tau}$ with $\tau^{-1} =
\Gamma_{atom} + \Gamma_{BG}$, to extract the value of
$\Gamma_{atom}$. $\Gamma_{atom}$ is related to the energy-dependent
cross-section, $\sigma(E)$, and the relative velocity, $v$, as
$\Gamma_{atom}$ = $n_{atom}\left<\sigma(E)v\right>$, where
$\left<~\right>$ denotes thermal averaging. Hence, knowledge of the
densities and temperatures allows the determination of the
scattering rate constant, $K(T) = \left<\sigma(E)v\right>$, for
these collisions.

\begin{figure}
\resizebox{1\columnwidth}{!}{
    \includegraphics{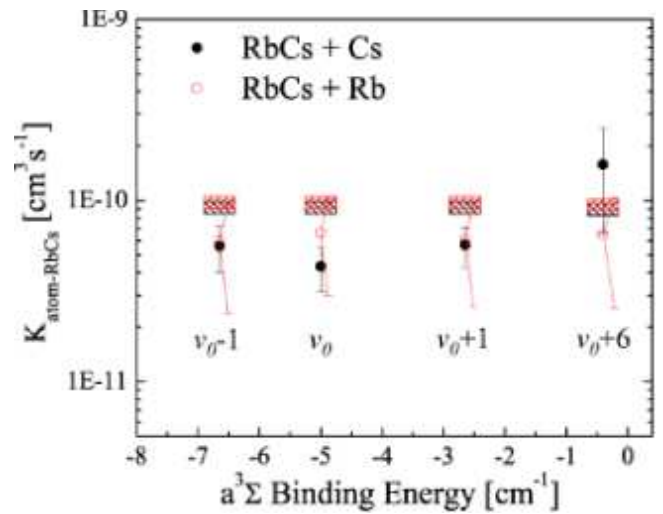}%
} \caption{(Color Online) . Molecular trap-loss scattering rate $K$
constant vs. binding energy $E_B$, for molecules in specific
vibrational levels of the a$^3\Sigma^+$ state. Vibrational state
label appears below each data point. Despite more than an order of
magnitude difference in binding energy, the scattering rates appear
to be identical within experimental precision. The error bars on
each point are the results of the uncertainties in the density and
lifetime measurements. Note that the Rb data's error bars have been
angled for clarity. The black cross-hatched box and red hatched box
are the prediction of the simple model described in the text for
collisions with Cs and Rb, respectively. The width of the boxes is
due to uncertainty in the collision temperature.
\label{AtomCollisionSummary}}
\end{figure}

Because the initial PA process populates several vibrational states
in the a$^3\Sigma^+$ state, we utilize the state-selectivity
provided by the REMPI detection to measure data similar to that in
Fig. \ref{LifetimeData} for a range of vibrational states and
therefore, a range of binding energies, $E_B \approx$ -0.5 cm$^{-1}$
to -7 cm$^{-1}$. The results of the collision measurements are
summarized in Fig. \ref{AtomCollisionSummary} as a function of $E_B$
(measured relative to the a$^3\Sigma^+$ asymptote). The measured
rate constants are identical within experimental precision, despite
over an order of magnitude of variation in $E_B$. Since the molecule
size and vibrational energy spacing both change substantially over
this range of $E_B$, the lack of dependence of the scattering rate
on molecular vibration hints at a unitarity limited process, where
the details of the short-range interaction potential are
unimportant. This view is supported by the agreement of the data
with the results of a simple model of the collision process, shown
in Fig. \ref{AtomCollisionSummary} as hatched boxes. This model
\cite{orzel_spin_1999}, which is detailed below, simply assumes that
any collision which penetrates to short-range results in an
inelastic trap loss event. Thus, it represents a more accurate
estimate of the upper-bound of an inelastic process than the usual
Langevin (unitarity-limited) cross-section
\cite{julienne_cold_1991}, which yields a scattering rate constant
that is roughly two to three times larger \footnote{The Langevin
scattering rate is given as $K_{Langevin} = \left<\frac{v
\pi}{k^2}\sum_{\ell}^{\ell^{max}}(2\ell + 1)\right> =
\left<\frac{\hbar\pi}{\mu k}(\ell^{max} + 1)^2\right>$, where
$\ell^{max}$ is defined as the value of $\ell$ for which the
centifugal barrier of the collision potential equals the kinetic
energy of the colliding particles.}.

For two colliding particles, the energy-dependent scattering
cross-section for the $\ell$th partial wave, with projection $m$,
from state $i$ to state $f$ in all outgoing partial waves $\ell',m'$
is generally written as
\begin{equation}
\sigma_{\ell,m}(E,i\rightarrow f) =
\frac{\pi}{k^2}\sum_{\ell',m'}\left|T_{\ell,m,\ell',m'}(E,i\rightarrow
f)\right|^2.\label{GeneralSigma}
\end{equation}
Here $T_{\ell,m,\ell',m'}(E,i\rightarrow f)$ is the `T-matrix',
whose elements represent the probability amplitude for a transition
from the incoming spherical wave, $\Psi_{i,\ell,m}$, to the outgoing
wave, $\Psi_{f,\ell',m'}$ and $k = \sqrt{2\mu E/\hbar^2}$ is the
magnitude of the wave-vector at collision energy E. Since our
experiments are sensitive to the total cross-section for collisions
that remove molecules from the trap, we must sum over all final
states $f$ that lead to trap loss, in addition to the normal sum
over $\ell,m$. If we assume that any collision that penetrates to
short-range is inelastic with unit probability, we can re-write Eq.
\eqref{GeneralSigma} as
\begin{equation}
\sigma(E,i) = \sum_{f,\ell,m}\sigma_{\ell,m}(E,i\rightarrow f) =
\sum_{\ell}\frac{\pi}{k^2}(2\ell+1)P_T(E,\ell),\label{SigmaTotal}
\end{equation}
where $P_T(E,\ell)$ is simply the probability of transmission to
short range. We calculate $P_T(E,\ell)$ by numerically solving the
Schr\"{o}dinger equation for the potential $V(r,\ell) =
\frac{\hbar^2\ell(\ell+1)}{2\mu r^2} - \frac{C_6}{r^6}$, and
assuming that any flux that does not reflect off the potential is
transmitted to short-range and then completely lost to inelastic
processes. This simple technique is applicable to any
highly-inelastic process; it requires only the knowledge of the
long-range part of the scattering potential, which in our case is
given entirely by the value of the van der Waals coefficient, $C_6$,
and reduced collision mass, $\mu$.

\begin{table}
\caption{Calculated $C_6$ coefficients for a$^3\Sigma^+$ RbCs($v$)
colliding with various partners given in atomic
units.}\label{C6Table}
\begin{tabular}{l|c|c|c|c}
  Collision Type & ($v_0 - 1$) & ($v_0$) & ($v_0 + 2$) & ($v_0 + 6$) \\
  \hline
  RbCs($v$) + RbCs($v$) & 65745 & 65086 & 64310 & 61291\\
  Rb + RbCs & 16991 & 16920 & 16869 & 15960\\
  Cs + RbCs & 19688 & 19604 & 19541 & 18482 \\
  \end{tabular}
\end{table}

In general, the $C_6$ constant for two colliding particles is given
by the integral over imaginary frequency of the product of the
particles' dynamic polarizabilities
\cite{derevianko_high-precision_1999}. Since the dynamic
polarizability of RbCs was calculated in Ref.
\cite{kotochigova_controlling_2006} and the atomic values are
well-known, the $C_6$ constant as a function of vibrational level is
straightforward to calculate. These are shown for reference in Tab.
\ref{C6Table}.

\begin{figure}
\resizebox{1\columnwidth}{!}{
    \includegraphics{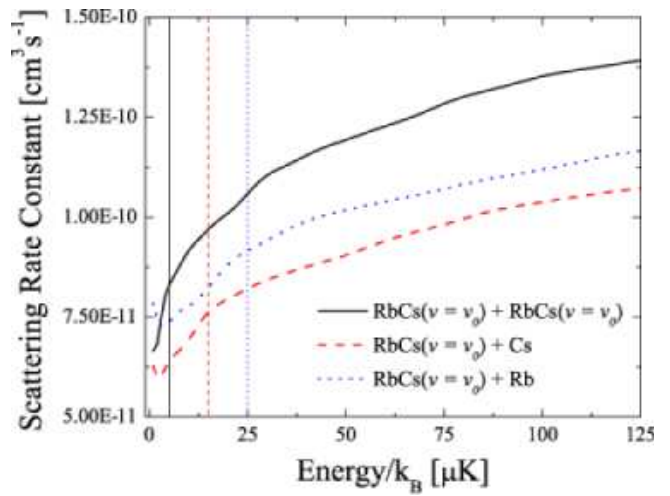}%
} \caption{(Color Online) Numerically calculated scattering rate
constant vs. energy for the three types of collisions. The p-wave
barrier height for each collision is represented by the vertical
lines with the same plot-style (color). \label{NumCalcRate}}
\end{figure}

Using these values, $P_T$, $\sigma$, and $K$ are calculated. The
results are shown in Fig. \ref{NumCalcRate}, where the scattering
rate, $K = \left<\sigma(E,i) v\right>$, is plotted versus collision
energy for the three classes of collisions. Note that the average
center-of-mass frame collision velocity is given as $\left<v\right>
= \sqrt{\frac{8k_bT_1}{\pi m_1} + \frac{8k_bT_2}{\pi m_2}} =
\sqrt{\frac{8k_bT_{\mu}}{\pi\mu}}$, where $T_i$ ($T_{\mu}$) is the
laboratory frame (center-of-mass frame) temperature and $m_i$ is the
mass of the colliding particles. Since the $\ell$ = 1 (p-wave)
barrier heights lie at $E/k_B \approx$ 25 $\mu$K, 15 $\mu$K, and 5
$\mu$K for RbCs colliding with Rb, Cs, and RbCs, respectively, the
collision rates fall to the values given solely by their s-wave
contribution at the lowest energies on the graph. Interestingly,
quantum reflection from the $\ell$ = 0 potential is found to scale
linearly with $k$ as $E\rightarrow0$, reproducing the Wigner
threshold law \cite{wigner_behavior_1948} for low temperature
inelastic scattering of $\sigma \propto 1/k$
\cite{balakrishnan_threshold_1997}. Thus the calculated scattering
rate remains finite at zero temperature, despite the fact that the
unitarity limited scattering rate scales as $v^{-1}$. From the
predicted molecule-molecule scattering rate, we expect an initial
two-body loss rate of $\frac{\beta}{V}N_{RbCs} = K n_{RbCs} \approx
0.1$ Hz. Given that $\Gamma_{BG} \approx 2$ Hz, our measurement of
background gas limited decay for the pure RbCs sample is consistent
with the predicted scattering rate.

In conclusion, we have demonstrated the trapping of heteronuclear
RbCs molecules in vibrationally excited levels of the a$^3\Sigma^+$
electronic ground state. Observations of the molecules in the trap
have revealed strong inelastic collisions, presumably due to a
combination of vibrational, rotational and hyperfine quenching. A
simple, extendable model, which relies only on the knowledge of the
van der Waals coefficient, accurately reproduces the observed rates
by assuming that all short-range collisions are inelastic for all
observed vibrational levels and both colliding atomic species.

We are currently working towards transferring these trapped
molecules into their absolute ground state, via the transfer scheme
previously demonstrated in our lab \cite{sage_optical_2005}. It
appears that with minimal improvements, \textit{e.g.} implementing
an adiabatic transfer \cite{bergmann_coherent_1998} instead of the
stimulated emission pumping used in \cite{sage_optical_2005}, a
sample of $>10^4$ absolute ground state molecules X$^1\Sigma(v = 0)$
at a temperature of 20 $\mu$K and density of $\geq10^{9}$ cm$^{-3}$
can be created. From calculations of the adiabatic transfer with
available laser powers, we estimate the transfer process can take
place in $\sim$100 $\mu$s; thus, we anticipate negligible loss of
population due to the inelastic collisions studied in the present
work. In fact, it appears these inelastic collisions could serve a
useful purpose. If Cs atoms are deliberately loaded into the
lattice, any molecule not in the absolute ground state will quickly
($\sim$100 ms) be removed from the trap by these inelastic
collisions. By contrast, X$^1\Sigma(v = 0)$ molecules, which cannot
undergo inelastic collisions with Cs atoms, are unaffected by the
presence of the atoms -- even ground state molecules can have
inelastic collisions with Rb atoms via the energetically allowed
substitution reaction RbCs + Rb $\rightarrow$ Rb$_2$ + Cs. After the
excited state molecules have been removed, resonant `push'-beams can
eject the remaining atoms from the trap, leaving behind a pure
sample of ground state molecules.

\bibliography{TripletCollisionsPaperBib}
\end{document}